\documentclass[doublecol]{epl2}

\title{Generalization of SUSY Intertwining Relations: New Exact Solutions of Fokker-Planck Equation}
\author{M.V. Iof\/fe\inst{1}\footnote{E-mail: m.ioffe@spbu.ru}
\and D.N. Nishnianidze\inst{2}\footnote{E-mail: cutaisi@yahoo.com}
}

\institute{
  \inst{1} Saint Petersburg State University, 7/9 Universitetskaya nab., St.Petersburg, 199034 Russia\\
  \inst{2} Akaki Tsereteli State University, 4600 Kutaisi, Republic of Georgia
}

\abstract{
It is commonly known that the Fokker-Planck equation is exactly solvable only for some particular systems, usually with time-independent drift coefficients. To extend the class of solvable problems, we use the intertwining relations of SUSY Quantum Mechanics but in new - asymmetric - form. It turns out that this form is just useful for solution of Fokker-Planck equation. As usual, intertwining provides a partnership between two different systems both described by Fokker-Planck equation. Due to the use of an asymmetric kind of intertwining relations with a suitable ansatz, we managed to obtain a new class of analytically solvable models. What is important, this approach allows us to deal with the drift coefficients depending on both variables, $x,$ and $t.$ An illustrating example of the proposed construction is given explicitly.
}

\begin{document}

\maketitle

\newcommand{\be}{\begin{equation}}
\newcommand{\ee}{\end{equation}}
\newcommand{\ba}{\begin{eqnarray}}
\newcommand{\ea}{\end{eqnarray}}
\renewcommand\thefootnote{\alph{footnote}}

\section{1. Introduction}

The Fokker-Planck equation is well known as one of the most widely used equations not only in Physics but also in Chemistry, Biology, Circuit Theory, Finances, etc.
The original paper of Fokker \cite{fokker} was devoted to study of Brownian motion of particles in a radiation field, and in the paper of Planck \cite{planck}, it was developed
to a complete theory of fluctuations\footnote{Sometimes, the terms Kolmogorov's Equation and Smoluchowski's Equation are also used.}.
Usually, the study of macroscopic qualitative changes of system is associated with random fluctuations which are most important in such changes of regime.
Briefly speaking, the Fokker-Planck equation describes those fluctuations in the system which are produced by many very small but unpredictable disturbances. The Brownian
motion problem is just a most well known example. Practically, as a rule, this equation allows to look for the probability distributions (for sure, positive valued) for some characteristic of the system. The literature on this theme is very rich, and it includes not only a lot of journal papers, but also several books, such as \cite{risken}, \cite{gardiner} \cite{kampen}.

Because of such importance of Fokker-Planck equation in a lot of branches of science, many approaches were used to find and investigate its solutions for different systems both in one and in several space dimensions. We may mention here: the separation of variables (eigenfunction method), stationary solutions, different kinds of boundary and initial conditions, variational method, numerical integration (see details in the same references \cite{risken}, \cite{gardiner}, \cite{kampen}). In this context, the supersymmetrical approach occupies an important place among others. This method appeared as a natural generalization of the modern supersymmetrical approach in the standard Quantum Mechanics (SUSY QM), which gave a new impetus to development of Quantum Mechanics and became very popular during last decades (see the wide review-like literature \cite{gendenstein}, \cite{cooper}, \cite{junker}, \cite{bagchi}, \cite{fernandez}, \cite{AI}). Historically, the SUSY QM has well known ancestors: the Factorization Method of Schr\"odinger \cite{hull} in Quantum Mechanics and the Darboux transformations \cite{matveev}, \cite{leble} for Sturm-Liouville equation in Mathematical Physics (see, for example, \cite{ABI}, \cite{ABEI}).
Originally \cite{witten}, the SUSY method establishes the symmetry between a pair of quantum non-relativistic systems described by the stationary Schr\"odinger equations. The so-called intertwining relations are the main ingredients of the SUSY method: two partner Hamiltonians  are intertwined by some differential operators (supercharges) leading to interconnection of spectra and wave functions of the partner systems \cite{cooper}, \cite{bagchi}, \cite{AI}. This approach was generalized in different directions, such as:
multidimensional systems \cite{AI}, \cite{ioffe}, \cite{ioffe-2}, systems with matrix potentials \cite{neelov}, exactly and quasi-exactly solvable models \cite{morse}, \cite{valinevich}, parasupersymmetry \cite{para} and many others.

Among a variety of generalizations of SUSY method, one is of special interest in the context of the present paper. We mean generalization to the case of nonstationary Schr\'odinger equation \cite{samsonov}, \cite{CIJN}, and to the case of Fokker-Planck equation which is close to it. The latter case was considered by different authors, for example: in \cite{bernstein}, \cite{CIJN} with usual first order supercharges, in \cite{schulze}, \cite{astarga} with higher order supercharges, in \cite{schulze-2-dim} with $(1+2)$ Fokker-Planck equation, in \cite{sasaki} with quasi-exactly solvable models. In the present paper, we shall develop further the supersymmetrical approach by means of new - asymmetric - form of intertwining relations, where instead of a pair of mutually conjugated intertwining operators (supercharges), two pairs are used. Thus, the intertwining relations become asymmetric, and a variety of their possible solutions become much more wide. This kind of generalization was used in \cite{shemyakova-1}, \cite{shemyakova-2} within some mathematical problems, and very recently in \cite{graphene} in the framework of investigation of graphene-like materials for solution of $(1+1)$ massless Dirac equation.
The paper is organized as follows. In the next section, we formulate the asymmetric generalization of intertwining relations for the Fokker-Planck equations, and we shall solve these relations in the framework of some suitable anzatses. In particular, one of the partner potentials $V_2$ is taken constant, but even this simple choice provides a wide variety of nontrivial solutions for the second potential $V_1(x,t).$ In the third section, an illustrative example will demonstrate how this algorithm can be used.

\section{2. Asymmetric intertwining of Fokker-Planck operators}

We start by definition of the $(1+1)-$dimensional Fokker-Planck (FP) operator with a constant diffusion coefficient $D=1$ and a drift coefficient $U(x,t),$ depending on both variables:
\ba
F[U]=&-&\partial_t+\partial^2 +\partial U^{\prime}(x,t)= \nonumber\\
       &-&\partial_t+\partial^2 + U^{\prime}(x,t)\partial + U^{\prime\prime}(x,t);\,\, \partial\equiv \frac{\partial}{\partial x}, \label{F}
\ea
where prime means derivative over $x.$ Solution $P(x,t)$ of the Fokker-Planck equation
\be
F[U]P(x,t)=0
\label{FP}
\ee
must be a real and positive function: it represents the probability distribution $P(x,t).$ It is convenient to transform the operator (\ref{F}) and the equation (\ref{FP}) into the equivalent form of diffusion equation for the function $\Psi(x,t):$
\ba
&&D[V]\Psi(x,t)\equiv\left(-\partial_t+\partial_x^2-V(x,t)\right)\Psi(x,t)=0;\nonumber\\
&&\Psi(x,t)=\exp{(U(x,t)/2)}P(x,t),
\label{D}
\ea
where the "potential" $V(x,t)$ is expressed in terms of the drift coefficient $U(x,t)$ as:
\ba
V(x,t)&=&\frac{1}{4}(U^{\prime}(x,t))^2-\frac{1}{2}U^{\prime\prime}(x,t) -\frac{1}{2}\dot{U}(x,t)\;  \label{VvsU}\\
\dot U(x,t)&\equiv &\frac{\partial}{\partial t}U(x,t). \nonumber
\ea
Although the equation (\ref{D}) seems to be similar (up to replacing $t \to it$) to the non-stationary Schr\"odinger equation, there are significant differences between them. Namely, 1) the operator $D[V]$ in (\ref{D}) is not Hermitian, and 2) its zero mode $\Psi(x,t)$ has to be real and positive like the probability $P(x,t).$

Of course, it is possible to solve the FP equation (and the equivalent diffusion equation) analytically only in certain specific cases. In this sense, the situation is completely analogous to the  situation with non-stationary Schr\"odinger equation. The FP equation appears in a great variety of physical problems, and beyond Physics as well, therefore, each opportunity to enlarge the class of analytically solvable cases of (\ref{FP}) has to be investigated. Among methods of solution, one - namely supersymmetrical (SUSY) method - must be mentioned here. Earlier, the method of SUSY transformations was generalized from the spectral problem with stationary Schr\"odinger equation to the problem of solution for non-stationary one \cite{samsonov}, \cite{CIJN}. Due to similarity between diffusion and non-stationary equations, this idea was successfully realized for FP equation as well by means of supercharges of first and higher orders in derivatives \cite{bernstein}, \cite{CIJN}, \cite{schulze}, \cite{astarga}, \cite{schulze-2-dim}, \cite{sasaki}.

In the present paper, instead of the usual ones, we shall explore the asymmetric intertwining relations:
\be
D[V_1(x,t)]\, N(x,t) = M(x,t)\, D[V_2(x,t)],
\label{IR}
\ee
for a pair of operators $D[V],$ where, in contrast to standard SUSY Quantum Mechanics, operators $N$ and $M$ do not coincide \cite{shemyakova-1}, \cite{shemyakova-2},  \cite{graphene}. As in original SUSY Quantum Mechanics \cite{witten}, analogues of supercharges $N,\, M$ are differential operators of first order in derivatives:
\ba
N(x,t)&=&f_0(x,t)\partial_t + f_1(x,t)\partial + f_2(x,t);   \label{N}\\
M(x,t)&=&g_0(x,t)\partial_t + g_1(x,t)\partial + g_2(x,t),    \label{M}
\ea
with coefficient functions which must be found from the intertwining relations (\ref{IR}). If we shall manage with solution of intertwining relations (\ref{IR}), i.e. if we shall find analytically both potentials $V_{1,2}$ and all coefficients $f_i,\,g_i$ for $i=0,1,2,$ a partnership between the zero modes $\Psi_1(x,t)$ and $\Psi_2(x,t)$ of operators $D[V_1]$ and $D[V_2],$ correspondingly, will be established. Indeed, for the known function $\Psi_2(x,t)$ such that
\be
D[V_2(x,t)]\Psi_2(x,t)=0,
\label{D2}
\ee
one obtains from (\ref{IR}) that the function
\be
\Psi_1(x,t) = N(x,t)\Psi_2(x,t)
\label{D21}
\ee
is the zero mode of $D[V_1]:$
\be
D[V_1(x,t)]\Psi_1(x,t)=0.
\label{D1}
\ee
Thus, if the problem (\ref{D2}) with chosen (more or less simple) potential $V_2(x,t)$ is solved, we shall obtain solution of the problem (\ref{D1}) with (probably, much more complicate) potential $V_1(x,t).$

The next task, solution of asymmetric intertwining relations, can be performed by equating coefficients for different derivatives in the left and right parts of (\ref{IR}). The result is the following:
\ba
&&g_0(x,t) = f_0(x,t); \label{0}\\
&&g_1(x,t) = f_1(x,t)=f_1(x,t)-2f_0^{\prime}(x,t); \label{1}\\
&&g_2(x,t) = f_2(x,t)+2f_1^{\prime}(x,t); \label{2}\\
&&V_2(x,t) = V_1(x,t)-2f_1^{\prime}(x,t); \label{3}\\
&&g_1(x,t) V_2(x,t) = f_1(x,t)V_1(x,t) + \dot f_1(x,t)  \nonumber\\
&&-f_1^{\prime\prime}(x,t) -2f_2^{\prime}(x,t); \label{4} \\
&&g_2(x,t) + g_0 V_2(x,t) = f_0(x,t)V_1(x,t) + \dot f_0(x,t)  \nonumber\\
&&- f_0^{\prime\prime}(x,t) + f_2(x,t); \label{5}\\
&&g_0(x,t) \dot V_2(x,t) + g_1(x,t) V_2^{\prime}(x,t) + g_2(x,t) V_2(x,t) = \nonumber\\
&&=f_2(x,t)V_1(x,t) + \dot f_2(x,t) - f_2^{\prime\prime}(x,t). \label{6}
\ea
First of all, it follows from (\ref{1}) and (\ref{0}) that both $f_0$ and $g_0$ depend only on time. Then, one can extract $f_0(t)$ from both sides of intertwining (\ref{IR}) modifying potential $V_1(x,t):$
\ba
&&f_0(t)D\biggl[V_1(x,t)+\frac{\dot f_0(t)}{f_0(t)}\biggr] \cdot \biggl(\partial_t + \tilde f_1(x,t)\partial + \tilde f_2(x,t)\biggr) =\nonumber\\
&&=f_0(t)\biggl(\partial_t + \tilde f_1(x,t)\partial + \tilde g_2(x,t)\biggr) \cdot D[V_2(x,t)].
\label{extract}
\ea
For this reason, below we shall take $f_0=g_0=1$ from the very beginning\footnote{Strictly speaking, we ignored one more option when both $f_0$ and $g_0$ vanish. In such a case, the system (\ref{1})-(\ref{6}) can be solved but only with $g_1=f_1$ and $f_2=g_2,$ i.e. with $N(x,t)=M(x,t).$ Thus, this option corresponds to old symmetrical intertwining relations which were studied earlier (for example, see \cite{CIJN}) and are not the subject of the present paper.}.

All equations of the system (\ref{1})-(\ref{5}) can be solved explicitly, and all coefficients can be expressed in terms of one arbitrary function $f(x,t):$
\ba
&&g_1(x,t) = f_1(x,t) \equiv f^{\prime}(x,t);\label{110}\\
&&g_2(x,t)=f_2(x,t)-2f^{\prime\prime}(x,t); \label{11}\\
&&f_2(x,t)=\frac{1}{2}((f^{\prime}(x,t))^2-f^{\prime\prime}(x,t)+\dot f(x,t)); \label{220}\\
&&V_2(x,t)=V_1(x,t)-2f^{\prime\prime}(x,t), \label{22}
\ea
while the last equation (\ref{6}) takes the following form of a nonlinear relation between the potential $V_2(x,t)$ and function $f(x,t):$
\ba
&&\dot f_2(x,t)+2f^{\prime\prime}(x,t)f_2(x,t)-f_2^{\prime\prime}(x,t)= \nonumber\\
&&=f^{\prime}(x,t)V_2^{\prime}(x,t)+2f^{\prime\prime}(x,t)V_2(x,t)+\dot V_2(x,t).
\label{33}
\ea
Since this relation can not be resolved in a general form, we must try to find as many particular solutions as possible.

Now we shall study rather simple ansatz which helps us find a variety of such solutions.
The constant potential:
\be
V_2(x,t)= c/2,
\label{c}
\ee
is a most obvious choice, for which Eq.(\ref{33}) is simplified essentially, and it is solved by:
\be
f_2(x,t)= c/2.
\label{cc}
\ee
Fortunately, eq.(\ref{220})
\be
c=(f^{\prime}(x,t))^2-f^{\prime\prime}(x,t)+\dot f(x,t)
\label{ff}
\ee
can be linearized by the substitution
\be
f(x,t)=ct-\ln\omega(x,t),
\label{omega}
\ee
where $\omega(x,t)$ is an arbitrary positive solution of the heat equation:
\be
\dot\omega(x,t) = \omega^{\prime\prime}(x,t).
\label{heat}
\ee
A wide variety of solutions of (\ref{heat}) is known for different domains on the plane $(x, t)$ and different kinds of boundary and initial conditions (see, for example, a collection in sect. 1.1.1 of the handbook \cite{polyanin}). Meanwhile, calculations above, together with (\ref{22}), allow us to build the partner potential $V_1(x,t),$ also in terms of chosen solution $\omega(x,t):$
\ba
&&V_1(x,t)=V_2(x,t)+2f^{\prime\prime}(x,t)=\frac{c}{2}-2(\ln\omega(x,t))^{\prime\prime}=\nonumber\\
&&=\frac{c}{2}-2\frac{\dot\omega(x,t)}{\omega(x,t)}+2\frac{(\omega^{\prime}(x,t))^2}{\omega^2(x,t)}.
\label{V1}
\ea

The diffusion equation (\ref{D2}) with constant potential $V_2=c/2$ allows a lot of solutions $\Psi_2(x,t)$, and the choice depends on the boundary conditions in the specific situation (see sect. 1.1.3 also in \cite{polyanin}, where the solutions of homogeneous equation with constant potential are listed). After choosing the solution $\Psi_2(x,t),$ the partner "wave function" $\Psi_1(x,t)$ can be obtained according to eq.(\ref{D21}):
\be
\Psi_1(x,t) = \biggl(\partial_t + f^{\prime}(x,t)\partial + \frac{c}{2}\biggr)\Psi_2(x,t).
\label{D2121}
\ee

Analogously to the partnership (\ref{D2121}) between $\Psi_1(x,t)$ and $\Psi_2(x,t),$ it would be interesting to find the direct partner relations between the drift coefficients $U_1(x,t)$ and $U_2(x,t).$ To achieve this goal, first of all it is necessary to recover the drift coefficients $U(x,t)$ from the corresponding potentials $V(x,t)$ using the relation eq.(\ref{VvsU}). Here again, the logarithmic substitution is definitely useful:
\be
U(x,t) \equiv -2\ln(\rho(x,t)),
\label{UZ}
\ee
so that:
\be
\rho^{\prime\prime}(x,t)=-\dot\rho(x,t) +V(x,t)\rho(x,t).
\label{UV}
\ee
We are interested in solutions $\rho_1(x,t)$ and $\rho_2(x,t)$ for potentials given by expressions (\ref{c}) and (\ref{V1}), correspondingly. In particular, for the constant potential $V_2:$
\be
U_2(x,t)=-ct-2\ln{(v(x,t))}; \,\, v^{\prime\prime}(x,t) = - \dot v(x,t),
\label{rho}
\ee
where the specific choice for solution $v(x,t)$ again depends on the specific boundary and initial conditions.

An explicit determination of the drift coefficient $U_1(x,t)$ (or, corresponding $\rho_1(x,t)$ of (\ref{UZ})) directly from equation (\ref{UV}) with a coefficient function $V_1(x,t)$ given by (\ref{V1}) seems to be really difficult problem. But some indirect way can be used to overcome this problem by choosing some suitable anzats - an additional restriction on a class of functions $\rho_1(x,t).$ One such hint can be obtained from the conjugate intertwining relation:
\be
N^{\dag}\, D^{\dag}[V_1(x,t)] = D^{\dag}[V_2(x,t)]\, M^{\dag}(x,t),
\label{conjugate}
\ee
where, in particular,
\be
D^{\dag}[V]=\partial_t+\partial^2 - V(x,t).
\nonumber
%\label{dagger}
\ee
From eqs.(\ref{VvsU}), (\ref{UV}), (\ref{UZ}) it follows that $\rho_1(x,t)$ is just a zero mode of $D^{\dag}[V_1].$ The solutions $f_i, \, g_i,\, V_1,\, V_2$ found above provide that both intertwining (\ref{IR}) and (\ref{conjugate}) are fulfilled as an operator relations. Acting by this operator relation on the function $\rho_1(x,t),$ we obtain from (\ref{conjugate}) that it is also a zero mode of r.h.s.. Although the r.h.s. contains a product of two operators, let us restrict ourselves with those $\rho_1(x,t)$ that are zero modes of $M^{\dag}.$ Thus, we have a system of two equations to define $\rho_1(x,t):$
\ba
&&\dot\rho_1(x,t) + \rho_1^{\prime\prime}(x,t)-(\frac{c}{2}+2f^{\prime}(x,t))\rho_1(x,t)=0; \label{55}\\
&&\dot\rho_1(x,t) + f^{\prime}(x,t) \rho_1^{\prime}(x,t)  \nonumber\\
&&-(\frac{c}{2}+f^{\prime\prime}(x,t))\rho_1(x,t)=0. \label{66}
\ea
Summing both, one obtains rather compact expression for $\rho_1(x,t)$ in terms of function $f(x,t)=ct-\ln\omega :$
\ba
&&\rho_1(x,t)=\exp{(f(x,t))} F(x,t); \nonumber\\
&&F(x,t) \equiv \beta(t) + \gamma(t)\int \exp{(-f(x,t))}dx,
\label{exp}
\ea
with arbitrary functions $\beta(t), \gamma(t).$ In addition, this expression must satisfy one of eqs.(\ref{55}), (\ref{66}). For example, we will insert it into eq.(\ref{66}). Straightforward calculations using formulas for derivatives:
\ba
&&F^{\prime}(x,t) = \gamma(t)\exp{(-ct)}\omega (x,t);
%\label{useful1}
\nonumber\\
&&\dot F(x,t) = \dot\beta(t) + \exp{(-ct)}[(\dot\gamma(t)-c\gamma )\cdot \nonumber\\
&&\cdot\int dx\omega (x,t) +\gamma(t)(\omega^{\prime}(x,t)+\alpha(t))],
\nonumber
%\label{useful2}
\ea
with $\alpha(t)$ one more arbitrary function (below for simplicity it will be taken $\alpha(t)=0),$ lead to the following result:
\be
\rho_1(x,t)= \frac{\exp{(ct/2)}}{\omega(x,t)}\biggl(a+b\int dx\omega(x,t)\biggr).
\label{rho-1}
\ee
Here, $\beta(t), \gamma(t)$ are parameterized by arbitrary positive constants $a, b$, as: $\beta(t)=a\cdot\exp{(-ct/2)};\,\, \gamma(t)=b\cdot\exp{(+ct/2)}.$

Finally, we may establish partnership between the probability distributions $P_1(x,t)$ and $P_2(x,t)$ which are solutions of FP equations. It is performed by inverse transformation according to (\ref{D}), and due to relation (\ref{UZ}), the result is the following:
\be
P_1(x,t) =\rho_1^{-1}(x,t) \biggl(\partial_t + f^{\prime}(x,t)\partial + \frac{c}{2}\biggr)\biggl(\rho_2(x,t)P_2(x,t)\biggr),
\label{final}
\ee
where $\rho_2(x,t)=\exp(ct/2)v(x,t)$ (see eq.(\ref{rho})), and $\rho_1(x,t)$  is given by (\ref{rho-1}).

\section{3. Example}

To illustrate the general algorithm for partnership between a pair of Fokker-Planck (and, correspondingly, between a pair of diffusion) operators, we consider here the specific solutions of equations in the previous section. First of all, we still consider the choice (\ref{c}) for one of potential: $V_2(x,t)=c/2.$ According to (\ref{V1}), the partner potential $V_1(x,t)$ is determined by the function $f(x,t)$ which is given by (\ref{omega}) with condition (\ref{heat}). The very wide variety of solutions of (\ref{heat}) (see details in  sect. 1.1.1 of \cite{polyanin}) provides a wide class of partner potentials $V_1(x,t).$ Let's choose rather simple but nontrivial variant:
\be
\omega(x,t)=(x^2 + 2t) + B;\quad  f(x,t) = ct-\ln\omega(x,t),
\label{e-1}
\ee
where $B$ is an arbitrary non-negative constant. Then, the potential
\be
V_1(x,t)= \frac{c}{2} + 4\frac{(x^2-2t)-B}{(x^2+2t)^2+2B(x^2+2t)+B^2}
\label{e-2}
\ee
obviously does not allow separation of variables in (\ref{D1}), and therefore, it is not straightforwardly solvable. Nevertheless, according to our construction above the wave function $\Psi_1(x,t)$ can be found from the wave function $\Psi_2(x,t)$ by means of intertwining relations (\ref{IR}).

Thus, now we need the solution $\Psi_2(x,t)$ of eq.(\ref{D}) with constant potential (\ref{c}). It is necessary to specify the task by choosing both the domain of the problem in $(x,t)$--plane and corresponding initial and boundary conditions. Let us choose the domain as a halfplane $t\geq 0, x\in (-\infty,\, +\infty),$ the vanishing values of $\Psi_2$ at $x\to\pm\infty$ and the point-like initial condition at $t=0:$
%\be
$$\Psi_2(x, t=0)=\delta(x).$$
%\nonumber
%\label{e-3}
%\ee
After substitution $\Psi_2(x,t)\equiv \exp{(-ct/2)}\varphi(x,t),$ the eq.(\ref{D2}) takes the form of the Cauchy problem for the heat equation for $\varphi(x,t)$ with the same boundary and initial conditions described above. The general expression for the solution with initial condition $\Psi_2(x, t=0)$ is \cite{polyanin}:
\ba
\Psi_2(x,t)&=& \exp{(-ct/2)}\varphi(x,t)=
%\label{e-4}
\nonumber \\
& &\frac{\exp{(-ct/2)}}{2\sqrt{\pi t}} \int^{+\infty}_{-\infty}dy \exp{(-\frac{(x-y)^2}{4t})}\varphi(y,t=0).
\nonumber
\ea
In our case, it takes the form:
%of fundamental solution for the heat operator
%\be
%(\partial_t-\partial^2_x)G(x,\ksi,t)=\delta(x-\ksi),
$$\Psi_2(x,t)=\frac{\exp{(-ct/2)}}{2\sqrt{\pi t}} \exp{(-\frac{x^2}{4t})}.$$
%\nonumber
%\label{e-5}
%\ee
%and in our case:
%\be
%\Psi_2(x,t) = \frac{Q}{2\sqrt{\pi t}} \exp{(-\frac{(x-\ksi)^2}{4t})}=Q
%\label{e-5}
%\ee

According to intertwining relations (\ref{IR}), the partner wave function $\Psi_1(x, t)$ is obtained by action of the operator $N(x,t)$ (see (\ref{D2121})):
\ba
&&\Psi_1(x,t)=(f^{\prime}(x,t)\partial +\partial^2)\Psi_2(x,t)= \nonumber\\
&&=[ \frac{x^2-2t}{4t^2} + \frac{x^2}{t(x^2+2t+B)} ]\frac{\exp{(-ct/2)}}{2\sqrt{\pi t}}\exp{(-\frac{x^2}{4t})}.
\nonumber
%\label{e-6}
\ea
This is solution of eq.(\ref{D1}) with potential (\ref{e-2}) which non-trivially depends on both $x$ and $t.$

The partnerships between pairs $V_1$ with $V_2$ and $\Psi_1$ with $\Psi_2$ can be extended to a partnership between drift coefficients $U_1$ with $U_2.$ However, it is necessary to remember that the same potential $V(x,t)$ corresponds to a lot of functions $U(x,t)$ given by a variety of solutions of eq.(\ref{UV}). In particular, for the constant potential $V_2(x,t)=c/2$ one may choose in (\ref{rho}) such a solution $v(x,t)$ that the drift coefficient $U_2(x,t)$ does not depend on $t:$
%\be
$$v(x,t)=\exp{(\beta x-\beta^2t)};  \,\, U_2(x,t)=-2\beta x; \,\, \beta = \sqrt{c/2}.$$
%\nonumber
%\label{e-7}
%\ee
In its turn, the particular form of $U_1(x,t)$ can be calculated following the construction in the very end of the second section. Inserting the expression (\ref{e-1}) for $\omega(x,t)$ into eq.(\ref{rho-1}), we obtain:
\ba
&&U_1(x,t)=-2\ln\rho_1(x,t)=-ct-2\ln(x^2+2t+B)   \nonumber\\
&&-2\ln[a+b(\frac{1}{3}x^3+2tx+Bx)].
\nonumber
%\label{e-8}
\ea
In order to avoid non-physical (complex) values of $U_1(x,t),$ we may somehow restrict ourselves. For example, we may choose the vanishing free parameter $b=0,$ or another choice - to consider both partner problems with the drifts $U_1(x,t)$ and $U_2(x,t)$ on a half line $x\geq 0.$

\section{4. Conclusions}

Long time ago, analytical methods of SUSY Quantum Mechanics were already used successfully in some problems of Classical Physics, and in particular, in study the well known Fokker-Planck equation \cite{junker}, \cite{CIJN}, \cite{bernstein}, \cite{schulze}, \cite{astarga}, \cite{schulze-2-dim}, \cite{sasaki}. In the present paper, this approach is continued by using modified ("asymmetric") intertwining relations for the Fokker-Planck operators. This modification of intertwining does not work for the spectral problem with the stationary Schrodinger equation, but it turned out to be useful here to solve $(1+1)-$dimensional Fokker-Planck equation. In particular, it allows to deal with drift coefficients nontrivially depending both on coordinates and time when the standard separation of variables is not possible. As usual for SUSY approach, intertwining relations establish the partnership between a pair of systems with different "potentials" and drift coefficients. If the problem for one (more simple) of these systems is solvable, the solution for the partner (more complicate) system is given by intertwining. In the second section above, the first system was chosen with a constant potential $V_2,$ but its partner corresponds to a whole class of potentials $V_1(x,t)$ nontrivially depending on $x$ and $t.$ The specific solutions depend on the choice of boundary conditions in coordinate and on initial Cauchy conditions for $t=0.$ As far as we know, our algorithm with asymmetric intertwining is the first constructive method which provides analytical solutions of Fokker-Planck equation with both $x-$ and $t$-dependence of drift coefficients. The example of the third section illustrates how this scheme works for a specific system.

\section{Acknowledgments}

The work of M.V.I. was supported by RFBR Grant No. 18-02-00264-a.

\end{document}